\documentclass[sigconf, anonymous=false]{acmart}

\usepackage{arydshln} 

\AtBeginDocument{%
  }

\setcopyright{acmlicensed}
\copyrightyear{2025}
\acmYear{2025}
\acmDOI{XXXXXXX.XXXXXXX}
\begin{document}

\title{Generating Search Explanations using Large Language Models}


\author{Arif Laksito}
\affiliation{%
  \institution{School of Computer Science, University of Sheffield}
  \city{Sheffield}
  \country{United Kingdom}}
\email{alaksito1@sheffield.ac.uk}

\author{Mark Stevenson}
\affiliation{%
  \institution{School of Computer Science, University of Sheffield}
  \city{Sheffield}
  \country{United Kingdom}}
\email{mark.stevenson@sheffield.ac.uk}


\begin{abstract}
Aspect-oriented explanations in search results are typically concise text snippets placed alongside retrieved documents to serve as explanations that assist users in efficiently locating relevant information.
While Large Language Models (LLMs) have demonstrated exceptional performance for a range of problems, their potential to generate explanations for search results has not been explored.
This study addresses that gap by leveraging both encoder-decoder and decoder-only LLMs to generate explanations for search results.
The explanations generated are consistently more accurate and plausible explanations than those produced by a range of baseline models. 
\end{abstract}



\maketitle

\section{Introduction}
In search systems, users frequently submit under-specified queries with multiple potential interpretations of the user's intent~\cite{MacAvaney.etal2021, Santos.etal2015, Iwata.etal2012}. This ambiguity often results in a diverse range of search results, requiring users to sift through numerous documents to find relevant information.
Snippets were introduced to help users quickly assess the relevance of a document to their query~\cite{Tombros.Sanderson1998}. These typically include the document's title, URL, and a brief summary of its contents, usually consisting of two to three lines. 
A recent study indicated that while snippets can enhance user interaction with search systems they often fall short of clearly explaining the relevance of the query to the retrieved documents~\cite{Rahimi.etal2021, Yu.etal2022}. 
One possible approach to representing multiple query intents involves specifying distinct information types—here referred to as “aspects.” 
For instance, in response to the underspecified query “badminton,” relevant aspects might include \textit{rules}, \textit{organization}, or \textit{equipment}.
Large Language Models (LLMs) have recently been shown to be highly effective for a wide range of text generation tasks. However, the capability of these models to generate concise, aspect-oriented explanations for search results remains unexplored. 

\section{Approach}
Previous work on aspect explanation generation has relied on modified Transformer architectures, incorporating a query attention layer in the encoder and masking the query in the decoder; these models were trained from scratch without leveraging pretrained checkpoints~\cite{Rahimi.etal2021}. In this work, we utilize fine-tuning of LLMs for both encoder-decoder and decoder-only models. Specifically, we perform full fine-tuning on smaller encoder-decoder models to generate explanations text. For larger decoder-only models, we adopt QLoRA~\cite{Dettmers.etal2023}, a parameter-efficient fine-tuning method that significantly reduces memory and computational requirements while maintaining competitive performance. Notably, QLoRA enables fine-tuning of large models on a single GPU by combining 4-bit quantization and low-rank adaptation.

\begin{figure}[ht]
\centering
\includegraphics[width=\linewidth]{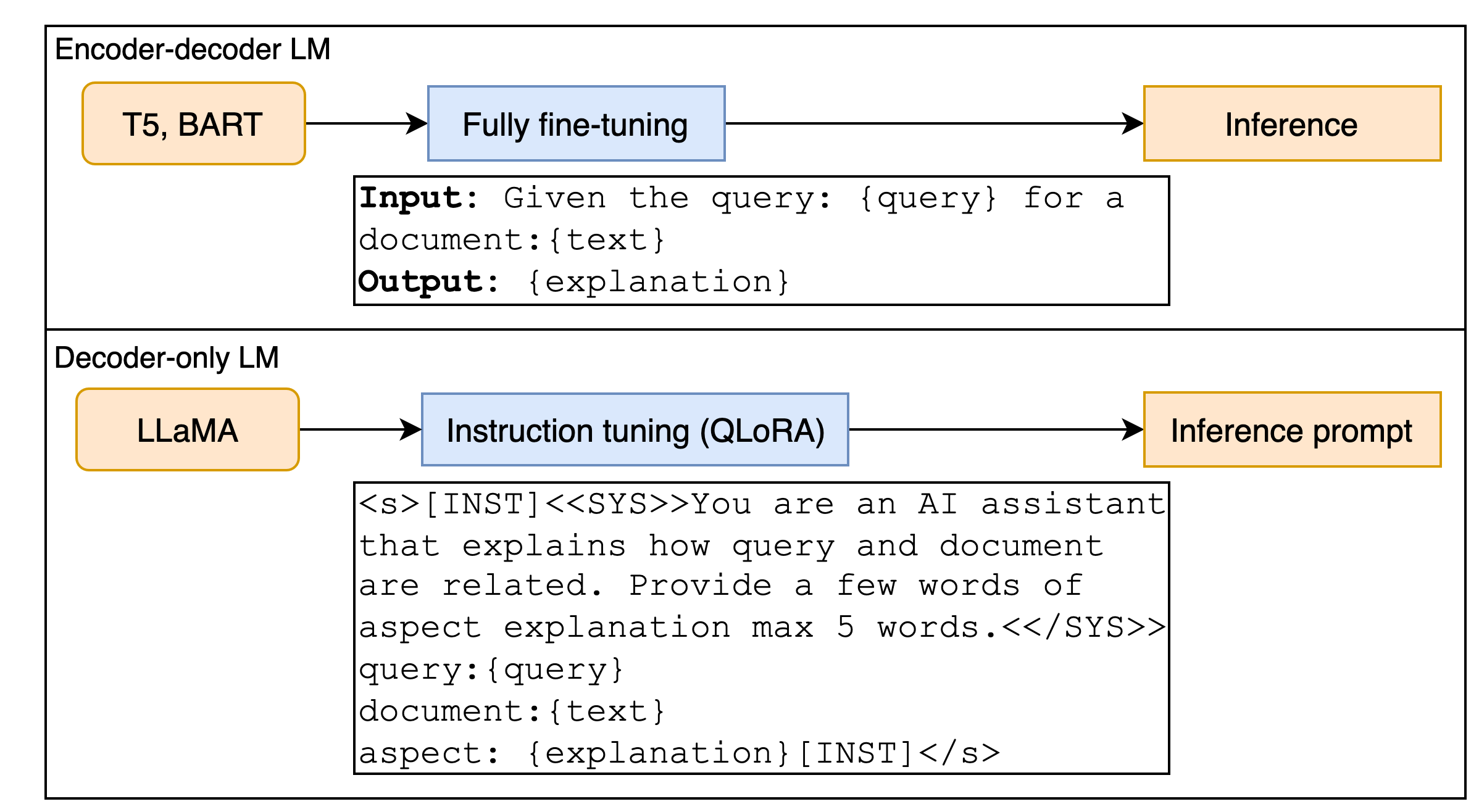}
\caption{Illustrates the different fine-tuning methods on encoder-decoder and decoder only language models} 
\label{fig1}
\Description{Fully fine-tuning and QLoRA fine-tuning.}
\end{figure}

Unlike previous studies~\cite{Rahimi.etal2021, Yu.etal2022} that rely on special token-based input formatting—such as inserting a [SEP] token to separate queries and documents—we employ a natural language input representation for training encoder-decoder models. For decoder-only models, we adopt an instruction-tuning framework where inputs are framed as natural prompts followed by expected outputs, aligning the task format with instruction-following behavior. An overview of the input-output structure used is illustrated in Figure~\ref{fig1}.

\begin{table*}[h]
\centering
\caption{Performance comparison of baseline models and our approaches. The highest score for each metric is highlighted in bold. Training time is reported for one epoch, and inference time is measured on 1,000 test samples.}
\label{table:rst-fine-tuning}
\begin{tabular}{lcccccrr}
\hline
 & \textbf{Architecture} & \textbf{Parameters} & \textbf{METEOR} & \textbf{ROUGE-1} & \textbf{BERTScore} & \textbf{Training time(s)} & \textbf{Inference time(s)} \\
\hline
Transformer & Encoder-decoder & 21M & 0.0747 & 0.1264 & 0.3057 & 393 & 60 \\
Bert2Bert & Encoder-decoder & 247M & 0.0846 & 0.1323 & 0.2970 & 2451 & 156\\
Bert2Gpt & Encoder-decoder & 262M & 0.1158 & 0.1917 & 0.3157 & 2586 & 163\\
0-shot LLaMA(v2) & Decoder-only & 13B & 0.0920 & 0.1145 & 0.1830 & - & 4,251\\
0-shot LLaMA(v3) & Decoder-only & 70B & 0.1215 & 0.1813 & 0.2920 & - & 15,974\\
\hline
FT BART & Encoder-decoder & 139M & 0.2331 & 0.3923 & 0.4771 & 1,589 & 148 \\ 
FT T5 & Encoder-decoder & 220M & 0.2723 & 0.4301 & 0.5202 & 3,232 & 153 \\
FT LLaMA(v2) & Decoder-only & 13B & 0.2759 & 0.3896 & 0.4362 & 9,566 & 3,506 \\ 
FT LLaMA(v3) & Decoder-only & 70B & \textbf{0.3222} & \textbf{0.4993} & \textbf{0.5652} & 141,211 & 27,292\\ 
\hline
\end{tabular}
\end{table*}

\section{Evaluation}
We constructed a dataset following the approach of treating Wikipedia article titles as queries and their section headings as aspect-based explanations~\cite{Rahimi.etal2021, Yu.etal2022}. Using the March 2024 English Wikipedia Dump, we selected articles with at least three relevant sections (128–512 tokens each). The dataset was split into training, development, and test sets by grouping queries and randomly assigning groups.\footnote{Full dataset construction details are available at:~\url{https://github.com/ariflaksito/en-wikisa}}

In explanation generation tasks, BLEU~\cite{Papineni.etal2002} and ROUGE~\cite{Lin2004} can be used to evaluate the overlap between the model output and the reference text by measuring word- and n-gram-level similarity. These metrics are widely adopted in machine translation and text summarization tasks due to their ability to quantify lexical overlap between generated and reference sequences.
However, METEOR~\cite{Banerjee.Lavie2005} has been shown to outperform BLEU by incorporating synonym matching, stemming, and paraphrase recognition, making it more sensitive to linguistic variation. 
In addition to these traditional metrics, we employed BERTScore, which leverages contextual embeddings from pretrained language models to compute token-level similarity based on meaning rather than surface form. 

Our approach was compared against several baseline models: 
\begin{enumerate}
    \item A standard encoder-decoder Transformer model~\cite{Vaswani.etal2017}, which utilizes the BERT tokenizer and vocabulary for input preprocessing.
    \item Bert2Bert and Bert2GPT configurations, which implement an encoder-decoder framework using BERT~\cite{Devlin.etal2019} as the encoder while differing in the decoder component—either leveraging BERT~\cite{Devlin.etal2019} or GPT-2~\cite{Radford.etal2019}, respectively.
    \item A zero-shot setup of LLaMA models~\cite{Touvron.etal2023} using prompt-based inference. 
\end{enumerate}

The encoder-decoder models described in points (1) and (2) above were trained from scratch, without initializing weights from any pretrained checkpoints. All encoder-decoder models, including both the baseline models  and the fine-tuned variants, were trained for a fixed number of 5 epochs to ensure consistency and comparability across experiments. Training was conducted under identical hyperparameter settings, with a learning rate of 1e-5 and a batch size of 8, using a single NVIDIA A100 80GB GPU. In contrast, due to the significantly larger parameter sizes and corresponding computational demands, both the LLaMA 13B and LLaMA 70B models were fine-tuned for only 1 epoch. This adjustment reflects practical limitations in training time and GPU memory, while still enabling meaningful model comparison.

\section{Results}

Table~\ref{table:rst-fine-tuning} presents the performance comparison between our fine-tuned models and several baselines. 
Overall, all fine-tuned models consistently outperformed the baselines across all evaluation metrics, underscoring the effectiveness of leveraging pretrained language models for this task.
Among all models, LLaMA v3 (70B) achieved the highest overall scores across all evaluation metrics, with a METEOR score of 0.3222, ROUGE-1 of 0.4993, and BERTScore of 0.5652, demonstrating the advantage of scaling up model size for generating consistent explanations. 

Zero-shot large decoder-only models, such as LLaMA, exhibited limited ability to generate concise text for this task, as shown in lower scores across all evaluation metrics compared to fine-tune approaches. This performance gap suggests that, despite their strong general language modeling capabilities, these models may lack the necessary task-specific conditioning to produce contextually appropriate explanations in the absence of supervised adaptation.

Notably, both fine-tuned BART and T5 models demonstrated strong performance in terms of both effectiveness and computational efficiency. These models outperformed all baselines by a substantial margin and achieved results that were not far behind the 13B and even 70B parameter LLaMA models. This highlights the efficiency and practicality of fine-tuning midsize encoder-decoder architectures, which can deliver competitive results while maintaining lower computational requirements for both training and inference. As such, fine-tuned BART and T5 models represent promising options for scenarios where computational resources are limited but high-quality generation is still required.

\section{Conclusion}
This work presents a comparative study on fine-tuning large language models (LLMs) for the task of generating aspect-based explanations using Wikipedia-derived data. We explore both encoder-decoder and decoder-only architectures. Experimental results demonstrate that larger models yield superior performance, with LLaMA v3 (70B) achieving the highest scores across all evaluation metrics. Nevertheless, smaller encoder-decoder models remain highly competitive, demonstrating robust results on all metrics while offering substantial improvements in training and inference efficiency.


\bibliographystyle{ACM-Reference-Format}
\bibliography{ref-sigir2025}


\end{document}